\documentclass{llncs}
\usepackage{makeidx}  % allows for indexgeneration
\usepackage{graphicx}
\usepackage{hyperref}
\begin{document}
\frontmatter          % for the preliminaries
\pagestyle{headings}  % switches on printing of running heads
\addtocmark{Hamiltonian Mechanics} % additional mark in the TOC

\mainmatter              % start of the contributions
\title{Methods to Expand Cell Signaling Models using Automated Reading and Model Checking}
%
% \titlerunning{Hamiltonian Mechanics}  % abbreviated title (for running head)
\author{Kai-Wen Liang\inst{1} \and Qinsi Wang\inst{1} \and Cheryl Telmer\inst{1} \and Divyaa Ravichandran\inst{1} \and Peter Spirtes\inst{1} \and Natasa Miskov-Zivanov \inst{2}}
\authorrunning{Kai-Wen Liang et al.} % abbreviated author list (for running head)
% \tocauthor{Ivar Ekeland, Roger Temam, Jeffrey Dean, David Grove,
%Craig Chambers, Kim B. Bruce, and Elisa Bertino}
%
\institute{Carnegie Mellon University, Pittsburgh PA 15213, USA,
\and
University of Pittsburgh, Pittsburgh, PA, 15213, USA}

\maketitle              % typeset the title of the contribution

\begin{abstract}
Biomedical research results are being published at a high rate, and with existing search engines, the vast amount of published work is usually easily accessible. However, reproducing published results, either experimental data or observations is often not viable. In this work, we propose a framework to overcome some of the issues of reproducing previous research, and to ensure re-usability of published information. We present here a framework that utilizes the results from state-of-the-art biomedical literature mining, biological system modeling and analysis techniques, and provides means to scientists to assemble and reason about information from voluminous, fragmented and sometimes inconsistent literature. The overall process of automated reading, assembly and reasoning can speed up discoveries from the order of decades to the order of hours or days. Our framework described here allows for rapidly conducting thousands of \textit{in silico} experiments that are designed as part of this process.

\end{abstract}

\section{Introduction}

Modeling, among many other advantages, facilitates explaining systems that we are studying, guides our data collection, illuminates core dynamics of systems, discovers new questions, or challenges existing theories\cite{epstein2008model}. However, the creation of models most often relies on intense human effort: model developers have to read hundreds of published papers and conduct numerous discussions with experts to understand the behavior of the system and to construct the model. This laborious process results in slow development of models, let alone validating the model and extending it with thousands of other possible component interactions that already exist in published literature. At the same time, research results are published at a high rate, and the published literature is voluminous, but often fragmented, and sometimes even inconsistent. There is a pressing need for automation of information extraction from literature, smart assembly into models, and model analysis, to enable researchers to re-use and reason about previously published work, in a comprehensive and timely manner.

In recent years, there has been an increasing effort to automate the process of explaining biological observations and answering biological questions. The goal of these efforts is to allow for rapid and accurate understanding of biological systems, treatment and prevention of diseases. To this end, several automated reading engines have been developed to extract interactions between biological entities from literature. These automated readers are capable of finding hundreds of thousands of such interactions from thousands of papers in a few hours\cite{Valenzuela+:2015aa}. However, in order to accurately and efficiently incorporate these pieces of knowledge into a model, we need a method to distinguish useful relationships from vast amounts of extracted information. The revised model often retains properties of the baseline model, but at the same time reflects new properties that the baseline model fails to satisfy, or suggests minimal interventions in the model that can lead to significant changes in outcomes.

To this end, the contributions of our work include: (i) Method to utilize previous research and published literature to validate existing knowledge about diseases, test hypotheses and raise new questions; (ii) Framework to rapidly conduct hundreds of \textit{in silico} experiments via stochastic simulation and statistical model checking; (iii) Pancreatic cancer microenvironment case study that demonstrates the framework's effectiveness.

\begin{figure}[ht]
    \centering
    \includegraphics[width=10cm]{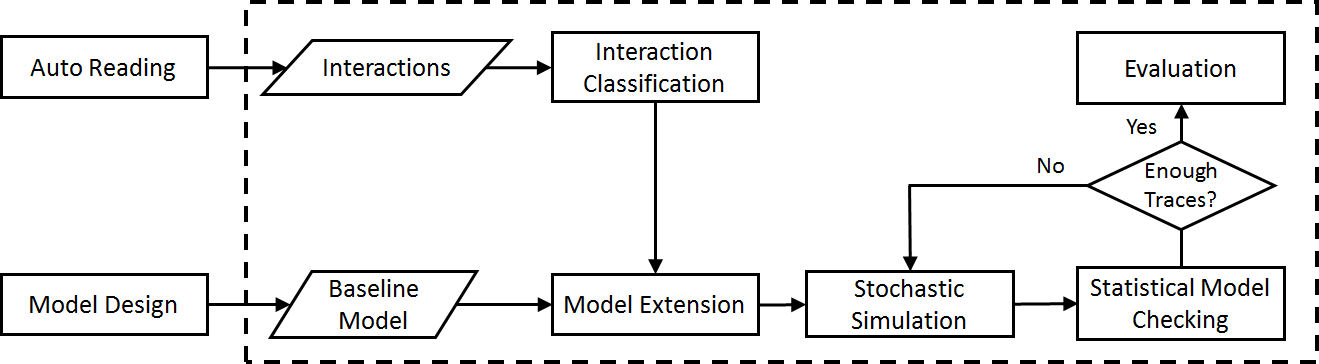}
    \caption{Steps of our model extension approach.}
    \label{fig:framework}
\end{figure}

Our framework is summarized in Fig. \ref{fig:framework}. The remainder of the paper is organized as follows. In Section 2 we provide details about the types of events extracted from literature. In Section 3, we outline methods to extend models. In Section 4, we describe model analysis methods. The results of applying our framework to pancreatic cancer microenvironment model are presented in Section 5. We discuss several important issues in Section 6 and conclude the paper with Section 7.

\section{Events in biomedical literature}

\begin{figure}[ht]
\centering
    \includegraphics[width=\textwidth]{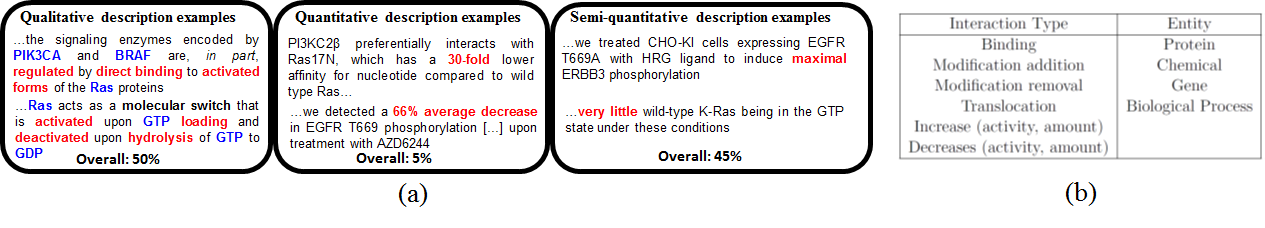}
    \caption{Reading output: (a) Examples of the three types of interactions found in papers and the average number of occurrences of each type in a sample paragraph set; (a) Types of interactions and their arguments (entities).}
    \label{fig:reader_exp}
\end{figure}

In this work, we focus on cellular pathways, that is, signal transduction, metabolic pathways and gene regulation. The literature that covers cellular pathways usually includes details such as molecular interactions, gene knock outs, inhibitors, stimulation with antigens. We conducted a brief exercise on a sample set of paragraphs from such published literature. The descriptions found in papers can be organized in three groups: qualitative, quantitative and semi-quantitative. Fig. \ref{fig:reader_exp}a shows examples of these three types of descriptions and the average number of occurrences for each type of interaction in the sample paragraph set. Automated reading engines\cite{Valenzuela+:2015aa} can extract events in the form of frames that contain an interaction with two entities (arguments). We list in Fig. \ref{fig:reader_exp}b the interaction and entity types that are recognized by reading engines and that we use in this work. Here, we represent each interaction as a pair $(u,v)$, where $u$ is the regulator and $v$ is the regulated element. For the first example sentence in Qualitative description  in Fig. \ref{fig:reader_exp}b, we can obtain two interaction pairs, $(Ras, PIK3CA)$, and $(Ras, BRAF)$.

\subsection{Baseline model type}
The interaction map of a model can be expressed as a \textit{directed} graph $\mathcal{G} = (V,E)$. The set of vertices, $V$, represents model elements, $v_i\in V$, $i=1..N$, where $N$ is the number of elements in the model. The set of edges, $E$, $(v_j,v_i) \in E$, represents causal interactions between elements, that is, relationships of type affects/is-affected-by. The polarity of interactions (positive or negative) is also included in the interaction map. 

In order to capture the type of information that most often occurs in published texts, as outlined in Fig. \ref{fig:reader_exp}(a), we are using logical modeling approach. In logical models of cellular signaling, each element from the interaction map $G$ has a corresponding Boolean variable $x_i\in \{0,1\}$. The update rule for a variable $x_i$ is a logic function of variables $x_j$'s, where each $x_j$ has a corresponding vertex $v_j\in V$, such that $(v_j,v_i) \in E$. That is, $f_i:\{0,1\}^{k_i} \to \{0,1\}$, where $k_i = |\{v_j: (v_j,v_i)\in E\}|$ is the in-degree of vertex $v_i$ . For a logical model with $n$ elements, there are $2^n$ possible configurations of variable values, and each configuration is called a state. The logical modeling approach works well with information extracted from text data, since the logical rules can be used to express the qualitative descriptions easily. For example, from the second example sentence in Qualitative description in Fig. \ref{fig:reader_exp}a, we could extract two interactions $(GTP, Ras)$ and $(!GDP, Ras)$, where '!' indicates negative regulation. We can implement all three elements, GTP, GDP, and Ras as Boolean variables, and write a logical rule for updating value of variable $Ras$ as, for example, $Ras = GDP\ and\ not\ GTP$.

\subsection{New interaction classification}
Often, the computational modelers start with a baseline model, and they add the information extracted from literature to the model. In order to add the extracted events, they first need to be classified according to their relationship to a given model. The output from reading engines can be related to the model in several ways: 

(i) \textit{Corroborations}: The interaction from reading output matches an interaction already in the model. An example of corroboration is shown with green arrow in Fig. \ref{fig:extension}a. 

(ii) \textit{Extensions}: The interaction from reading output is not found in the baseline model. An example of extension is shown with blue arrow in Fig. \ref{fig:extension}a. 

(iii) \textit{Contradictions}: The interaction from reading output suggests a different mechanism from the model (for example, activation vs. inhibition). An example is shown with red arrow in Fig. \ref{fig:extension}a. In this work, we study \textit{extensions only}, that is, new interactions that can be added to the model. Handling contradictions is part of our future work. 
\label{par:interaction_classification}

\section{Model extension}

\label{model_method}

\begin{figure}[ht]
    \centering
    \includegraphics[width=\textwidth]{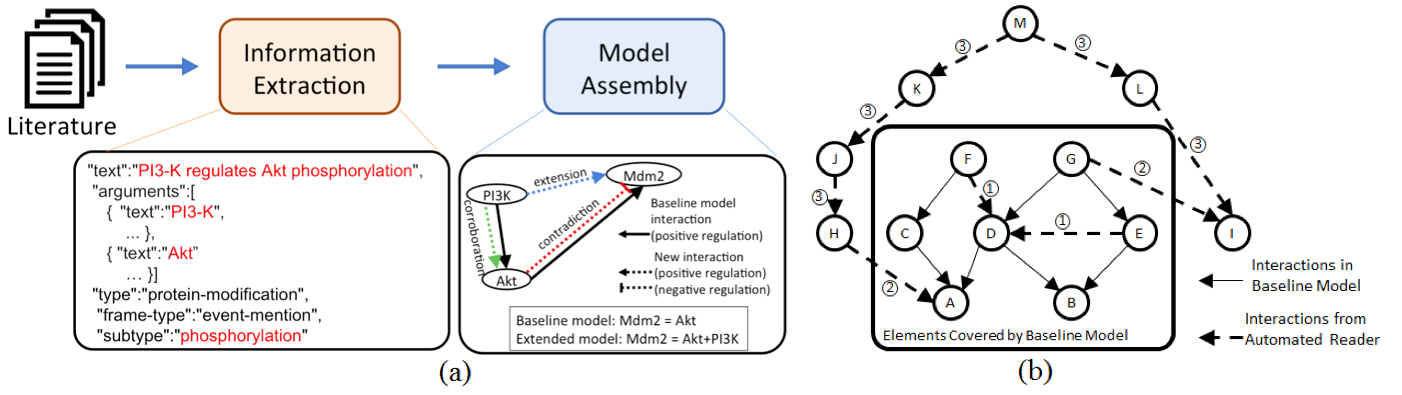}
    \caption{Relationship between reading output and model. (a) The literature reading-assembly flow with example. (b) Example baseline model (solid arrows) and new interactions extracted by automated reading (dashed arrows). The circled numbers represent classification described in Sec. \ref{par:interaction_classification}}
    \label{fig:extension}
\end{figure}

In Fig. \ref{fig:extension}b we show a toy example of model interaction map (solid arrows) and several extensions extracted by automated reading (dashed arrows). There are three kinds of model extensions (illustrated in Fig. \ref{fig:extension}b):

1.  Interactions where both elements are already in the model (edges $(E,D)$ and $(F,D)$ in Fig. \ref{fig:extension}b). This kind of extension usually has a direct influence on the behavior of the model: when adding a new interaction between elements in the model, we are creating a new pathway or generating feed-forward or feedback loops. These structural changes may lead to a significant difference in the regulatory behavior.

2. Interactions where only one element is in the baseline model (for edge $(H,A)$ in Fig. \ref{fig:extension}b the regulated element is in the baseline model, while the regulator is not; for edge $(G,I)$ the regulator is in the baseline model while the regulated element is not). In cases where the regulated element is not in the baseline model, the regulated element will just hang from a pathway without having direct influence on the model. On the other hand, in extensions where the regulator is outside the baseline model, the regulator can act as a new model input, allowing for additional network control.

3. Interactions consist of elements outside the baseline model (edges $(M,K)$, $(K,J)$). Such interactions alone do not affect the behavior of the model. However, when we are considering multiple extensions simultaneously, additional regulatory pathways may be constructed that will have effect on model behavior. The path $M\to K\to J\to H\to A$ in Fig. \ref{fig:extension}b is an example of newly formed pathway.

\begin{figure}[ht]
    \centering
    \includegraphics[width=\textwidth]{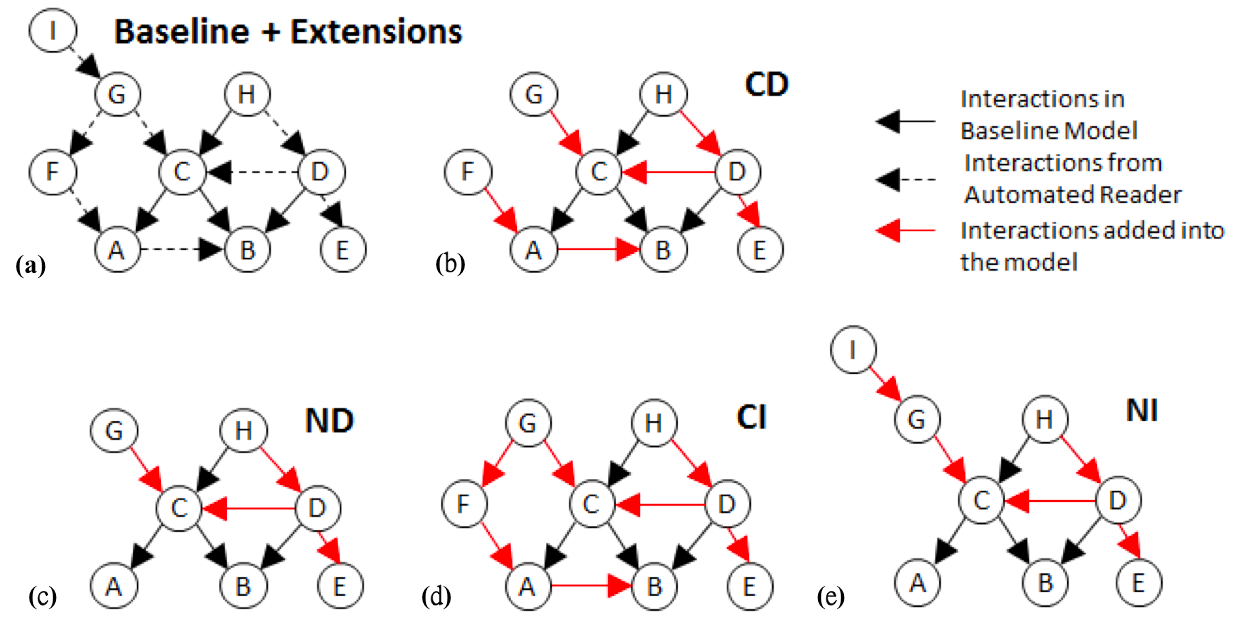}
    \caption{Results from different extension methods: (a) The baseline model and the extensions from automated reading; (b) The result from method CD with $n=1$; (c) The result from ND with $n=1$; (d) The result from CI with $n=1$; (e) The result from NI with $n=1$ and $m=1$.}
    \label{fig:ext_method}
\end{figure}

\subsection{Interaction map extension}

Each interaction of extension type can be regarded as a candidate new edge in the model's interaction map. Let $E^{\mathrm{ext}}$ be the set of interactions provided by reading. Suppose the baseline model is $\mathcal{G} = (V,E)$. Each new candidate model can be obtained by adding a group of selected edges $E^{\mathrm{new}} \in E^{\mathrm{ext}}$ and its corresponding elements, that is, $\mathcal{G}' = (V',E')$, where $E'=E \cup E^{\mathrm{new}}$. However, it is impossible to enumerate all configurations of whether or not to add a new edge, as the number of candidate models will become extremely large. For example, if there are $100$ new interactions extracted by the automated reader, there are $2^{100}$ possible extensions of the model. This number is impossible to handle, therefore, we need heuristic methods to search for suitable configurations of model extensions.

A possible way to tackle the issue of large number of model extension configurations, is to list the elements of interest in the baseline model, and include, as an extension, interactions that are related to those elements. The set of model elements of interest can be defined by user, depending on the questions asked or hypotheses tested. Still, the extension configurations need to be constructed in a systematic manner. Here we introduce the concept of 'layer', where layer $S_0$ is the set of elements of interest. The next layer, $S_1$, is the set of direct parents of elements in $S_0$, and in general, $S_i$ is the set of direct parents of elements in $S_{i-1}$. Elements in $S_1$ are direct regulators of $S_0$, and thus, the extensions including elements in $S_1$ may influence the model. Using this concept, we propose four different methods to create extension configurations. 

{\bf Cumulative parent set with direct extensions (CD)}: In this method, we define the number of layer, $n$, that we want to consider, and include all new interactions that affect any element from layer 
% Shouldn't the following be layer S0 up to layer Sn? Yes!
$0$ up to layer $n$. In other words, we add an extension $e=(u,v)$ to the model when at least one of the nodes  $u$ and $v$ is mentioned in layers $S_0$ to $S_n$. Fig. \ref{fig:ext_method}b demonstrates the result of this method where $n$ equals $1$. Starting from $S_0 = \{A,B,E\}$, we find its direct parents $S_1 = \{C,D\}$. The edges in the figure represent the union of layers $S_0$ and $S_1$. The advantage for this method is that it includes as many relevant extensions as possible within a certain distance from the elements of interest. However, due to the large number of elements added, the behavior of the model may become intractable within a few layers, that is, if the behavior deviates from what we expect, it is hard to pin-point the source of the change.

{\bf Non-cumulative parent-set with direct extensions (ND)}: This method can be used when we want to know the influence only from the $n^{th}$ layer. When creating each layer, we exclude the elements that are already mentioned in the previous layer, and repeat this process for $n$ times. As a result, from all elements in the ND set, layer $0$ can be reached within $n$ steps. We add extension $e=(u,v)$ to the model if and only if $u$ or $v$ is in the ND set. Fig. \ref{fig:ext_method}c is an example with $n=1$. After acquiring the layer $S_1$, we exclude the elements in the previous layer $S_0$, so we only include edges containing elements in $S_1$. Compared to the results of the CD method, ND method helps identify individual extension layers that may cause significant changes to the performance in different properties.

{\bf Cumulative parent-set with indirect extensions (CI)}: In the previous two methods, we find each layer only by looking for direct parents of previous layer, that is, the regulators that are already in the baseline model. In this method, we also look for indirect parents. In the example shown in Fig. \ref{fig:ext_method}d, we start from $S_0 = \{A,B,E\}$, which has as direct parents nodes $C$ and $D$, and as an indirect parent node $F$. Therefore, if we consider indirect parents, $S_1$ includes $\{C,D,F\}$. The result in Fig. \ref{fig:ext_method}d is obtained by adding edges including elements in $S_0$ or $S_1$ into the model. This method incorporates more elements into the model, allowing us to examine the behavior of the model including all edges within certain layers. It also includes pathways outside the baseline model more often then the other methods. However, just like the first method, the behavior here may become intractable when $n$ is large, especially when the network outside the baseline model is complicated.

{\bf Non-cumulative parent-set with indirect extensions (NI)}: This method is the combination of the two previous methods. The goal of this method is to provide information about influence on property values of $m$ layers containing indirect edges, starting from the $n^{th}$ layer. In other words, we first look at the $n^{th}$ layer using the ND method, and perform the operation of CI for $m$ times to find all the layers we are interested in. From Fig. \ref{fig:ext_method}e, we can see that using one ND step, we get the layer $S_1=\{C,D\}$. Using CI for another time, we have the set $S_{1,1} = \{G,H\}$. Adding elements mentioned in $S_1$ and $S_{1,1}$ results in the structure in \ref{fig:ext_method}e. This method can be more comprehensive than ND, giving us a more thorough understanding of the extensions. However, it could also suffer from the issue of being intractable if $m$ is large.

\subsection{Executable rule updating}

After choosing extension classification method and proper parameters for layer numbers, we create model extension sets. These sets extend the static interaction map of the model. Logical rules, on the other hand, allow for dynamic analysis of the model, as variable states change in time according to their update functions. Therefore, the set of logic update rules represents executable model. Incorporating new components into executable model rules can be done in several different ways. For example, if the original rule is $A = B or C$, and the extension interaction states that $D$ positively regulates $A$, then the new update rule for $A$ can be either $A = (B or C) and D$, or $A = B or C or D$. Other logic functions could be derived as well, but this largely depends on the information available in reading output about these interactions. Given that individual reading outputs only provide information of type 'participant a regulates participant b' (in our example, $D$ positively regulates $A$), and no additional information about interactions with other regulators is given (in our example, that would be combined regulations of $A$ by $B$, $C$ and $D$), we use two naive approaches, which is to add new elements to update rules using either $OR$ or $AND$ operation.

\section{Property testing}

After obtaining different extended models using the methods in Section \ref{model_method}, we evaluate the performance of each model by checking whether each extended model satisfies a set of biologically relevant properties. While simulations of logical models are known to be able to recapitulate certain experimental observations\cite{miskov2013duration}, verifying the results of the simulation against the properties manually is tedious and error-prone, especially when the number of models or properties becomes large. A feasible way to tackle this problem is to use formal methods. We use statistical model checking that combines simulation and property checking on simulation traces to compute the probability for satisfying each property. We elaborate each framework component in the following subsections.

\subsection{Stochastic simulation}
\label{sec:simulator}

The original logical model and all the extended model versions are simulated using stochastic method. We identify initial states for all model elements, ${\bf x}=(x_1,...,x_N)$, by assigning initial values to their corresponding Boolean variables ${\bf x}_0 = \{0,1\}^{n}$. Next, we use update rules to compute new variable values, that is, new states of all model elements. The simulator we use is publicly available \cite{miskov2014thimed,Kuo2015JAVA}. In the simulator, several different simulation schemes were designed to reflect different timing and element update approaches occurring in biological systems. The simulation scheme we use for this work is called \textit{Uniform Step-Based Random Sequential (USB-RandSeq)} 
% include citation for the simulator
. In each simulation \textit{step}, one model element is \textit{randomly} chosen, its update function is evaluated, and the value of its corresponding variable is updated. At the beginning of simulation the number of these \textit{sequential} steps is defined. In the case of \textit{uniform} update approach, all variables have the same probability of being chosen. The variable values in each step, starting from the initial state, ${\bf x}_0$, are recorded in a trace $\sigma = ({\bf x}_0,{\bf x}_1,\dots,{\bf x}_n)$. With the trace file at hand, we can use model checker to automatically verify whether or not the model meets several properties. Since the order of updating elements is random, when we run simulator to obtain multiple traces, the traces of variable values across different runs can vary.

\subsection{Statistical model checking}

The simulation of logical models is similar to discrete-time Markov chain, which means the verification problem is equivalent to computing the probability of whether a given temporal logic formula is satisfied by the system. One approach is to use numerical methods to compute the exact probability; however, this naive implementation suffers from the state explosion problem, and does not scale well to large-scale systems \cite{younes2002probabilistic}. Statistical model checking provides an excellent solution to this problem, by estimating the probability using simulation and thus, avoiding a full state space search. To verify a model via statistical model checking against interesting properties, we first need to encode each property into temporal logic formulae. Here we use Bounded Linear Temporal Logic (BLTL) \cite{jha2009bayesian}. BLTL is a variant of Linear Temporal Logic \cite{pnueli1977temporal}, where the future condition of certain logic expressions is encoded as a formula with a time bound (see the supplementary material (http://ppt.cc/XlWF7) for BLTL's formal syntax and semantics). To verify whether a model satisfies the properties, statistical model checking treats it as a statistical inference problem for the model executions generated using the randomized sampling. For a stochastic system, the probability $p$ that the system satisfies a property $\phi$ is unknown. Statistical model checking can handle two kinds of questions: (i) for a fixed $0 < \theta < 1$, determine whether $p\leq \theta$, and (ii) estimate the value of $p$. The first problem is solved using hypothesis testing methods, while the second is solved via estimation techniques. Statistical model checking assumes that, given a BLTL property $\phi$, the behavior of a system can be modeled as a Bernoulli random variable $M$ with parameter $p$, where $p$ is the probability of the system satisfying $\phi$. Statistical model checking first generates independent and identically distributed samples of $M$. Each sample $\sigma$ is then checked against the property $\phi$, and the yes/no answer corresponds to a 1/0 sample of the random variable $M$. The sample size does not need to be fixed, as the checking procedure will stop when it achieves the desired accuracy. This reduces the number of samples needed. The statistical model checking ha been applied in the past to the type of stochastic simulation that we use here, \cite{vardi1985automatic}.

\section{Results}

The system that we studied is pancreatic cancer microenvironment, including pancreatic cancer cells (PCCs) and pancreatic stellate cells (PSCs). We adopted here the model created by Wang et al.\cite{wang2015mpc}, which has three  major parts: (1) intracellular  signaling network of PCC; (2) intracellular signaling network of PSC; (3) network located in extracellular space of the microenvironment, which contains mainly ligands of the receptors. In this model, several cellular functions, such as autophagy, apoptosis, proliferation, migration, are also implemented as elements of the model, which enables modeling of the system's behavior that can result from turning various signaling components ON or OFF. In total, there are 30 variables encoding intracellular PCC elements and 3 variables encoding PCC cellular function. For PSC, there are 24 variables for intracellular elements and 4 variables for PSC cellular function. In extracellular microenvironment, there are 8 variables encoding extracellular signaling elements with 1 environment function variable. Accordingly, there are 70 variables in the model that have associated update functions used to compute next state of those model elements. The interaction rules of this model are summarized in Table 1 in the Supplementary material (http://ppt.cc/XlWF7).

The framework is implemented in Python. The simulator described in Section \ref{sec:simulator} is implemented in Java\cite{Kuo2015JAVA}. We use PRISM\cite{kwiatkowska2011prism} as our statistical model checker, which is a C++ tool for formal modeling and analysis of stochastic systems. Evaluating a model against one property, including running the simulations, takes about $10$ minutes on a regular laptop (1.3GHz dual-core Intel Core i5, 8GM LPDDR3 memory). The other components in the framework take less than $1$ minute. We used the REACH automated reading engine \cite{READINGOUTPUT} output produced from 13,000 papers in publicly available domain. This output consists of 500,000 event files, with 170,000 possible extensions of our model (other events are corroborations or contradictions).

To demonstrate how our framework works, we identified elements of interest in the model (which were suggested by cancer experts), and defined a set of relevant properties reflecting important biological truths that the PCC-PSC model should satisfy\cite{wang2016model}. In Table \ref{tab:properties}, we list $20$ properties that we tested using statistical model checking. There are five major functions or phenomena that we are interested in: (1) increased secretion of important growth factors; (2) over-expresion of oncoproteins in PCC and PSCs; (3) inhibition of tumor suppressors in PCCs; (4) cell functions of PCCs; (5) cell functions of PSCs.

\begin{table}[ht]
    \centering
    \includegraphics[width=\textwidth]{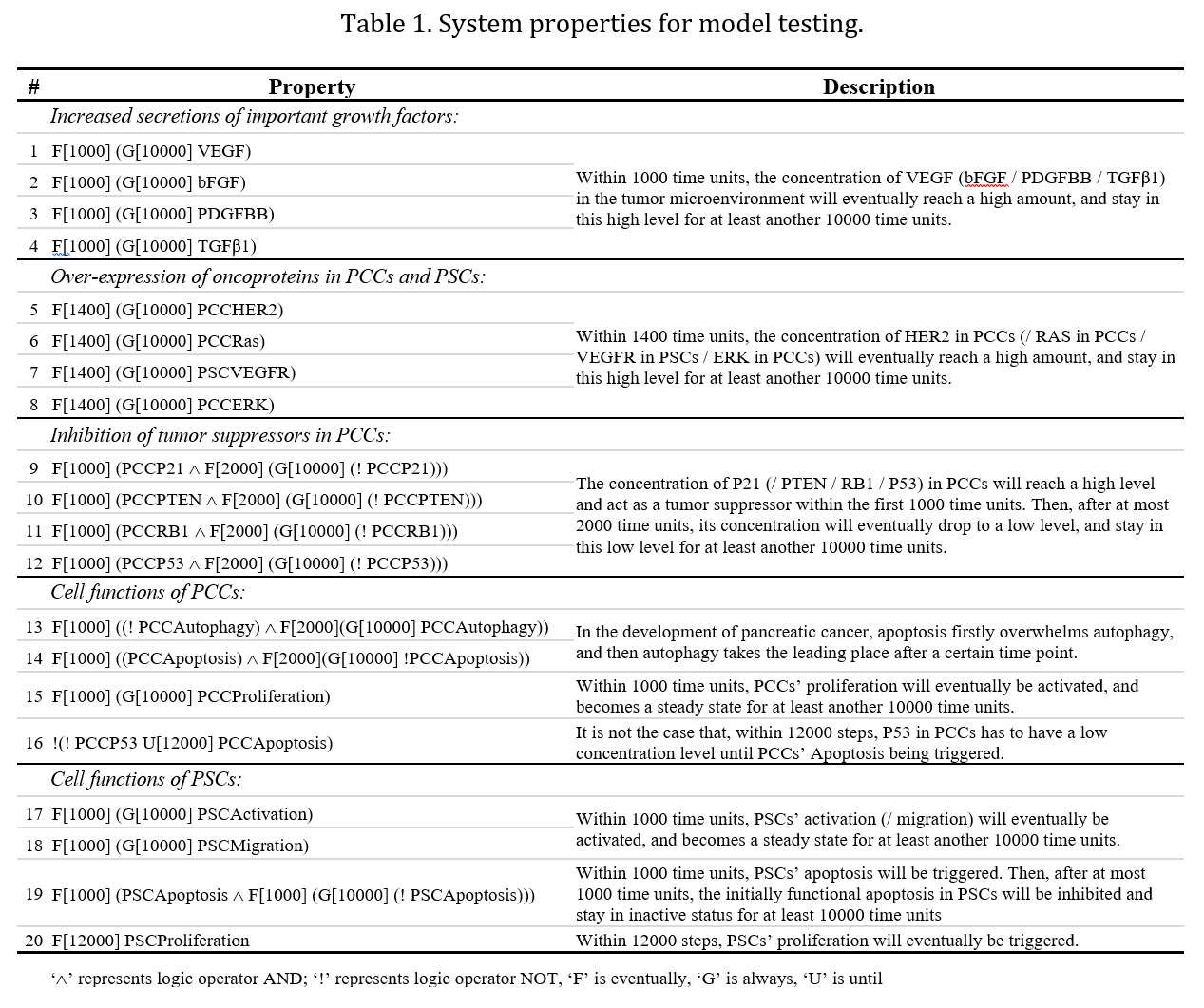}
    \label{tab:properties}
\end{table}

\subsection{Impact of proposed extension approaches on model}

The baseline model\cite{wang2015mpc} has 70 elements and 114 regulatory interactions. Although there are 170,000 model extensions produced by reading, many of them are repetitions, and some of the reading outputs were missing one of the interaction participants. Therefore, in this work we used overall 1232 different interactions from reading output, which could lead to $2^{1232}$ possible models. Studying all possible model versions is impractical, and therefore, we used the four extension methods described in Section \ref{model_method}, to generate $46$ different models. Using the CD method, we generated $2$ models by having $1$ or $2$ layers. For ND, the number of layers we considered varied between $1$ and $10$, which resulted in $10$ models. With CI, we used either $0, 1, 2$ or $3$ layers, which led to $4$ different models. Finally, for NI, we have $n$ ranges from $1$ layer to $10$ layer, and $m$ ranges from $1$ to $3$, resulting in $30$ models. We also test the model with all extensions being added to the baseline model.

\begin{figure}[ht]
    \centering
    \includegraphics[width=0.8\textwidth]{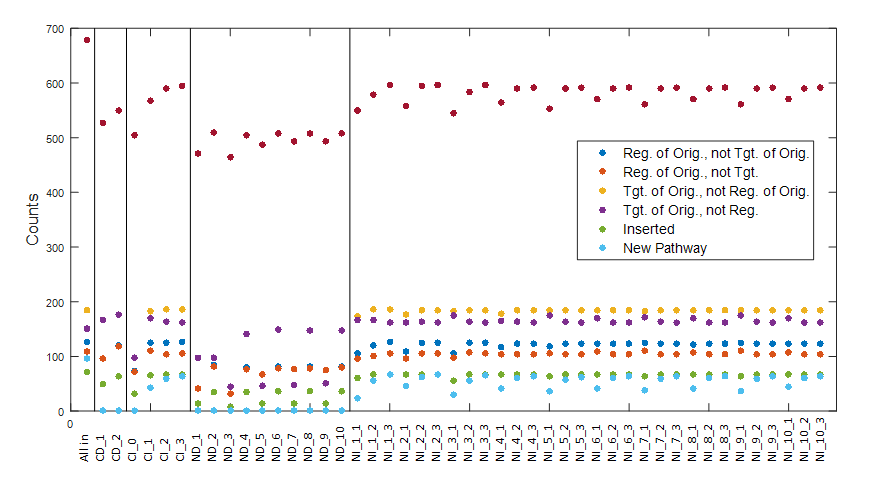}
    \caption{Counts for newly added elements with certain structure (\textbf{Reg}. - regulator, \textbf{Tgt}. - regulated element, \textbf{Orig}. - baseline model elements, \textbf{New} - newly added element). All models studied are listed on x-axis, and y-axis is the count of new elements having certain structure.}
    \label{fig:model_info}
\end{figure}

Fig. \ref{fig:model_info} summarizes results of our extension methods on 1232 interactions with respect to new node connections to the model: 

(i) number of new nodes regulating baseline model elements, not regulated by baseline model elements (dark blue); 

(ii) number of new nodes regulating baseline model elements, not regulated by any element, baseline or new (red); 

(iii) number of new nodes regulated by baseline model elements, not regulating any elements in the baseline model (yellow); 

(iv) number of new nodes regulated by baseline model elements, not regulating any element, baseline or new (purple); 

(v) number of new nodes inserted into existing pathway - new regulators of baseline model elements that are also regulated by baseline model elements (green); 

(vi) number of new nodes as intermediate elements of new pathways when multiple extensions are connected (light blue); 

(vii) total number of all elements used in the extension method (dark red). 

\begin{figure}[ht]
    \centering
    \includegraphics[width=\textwidth]{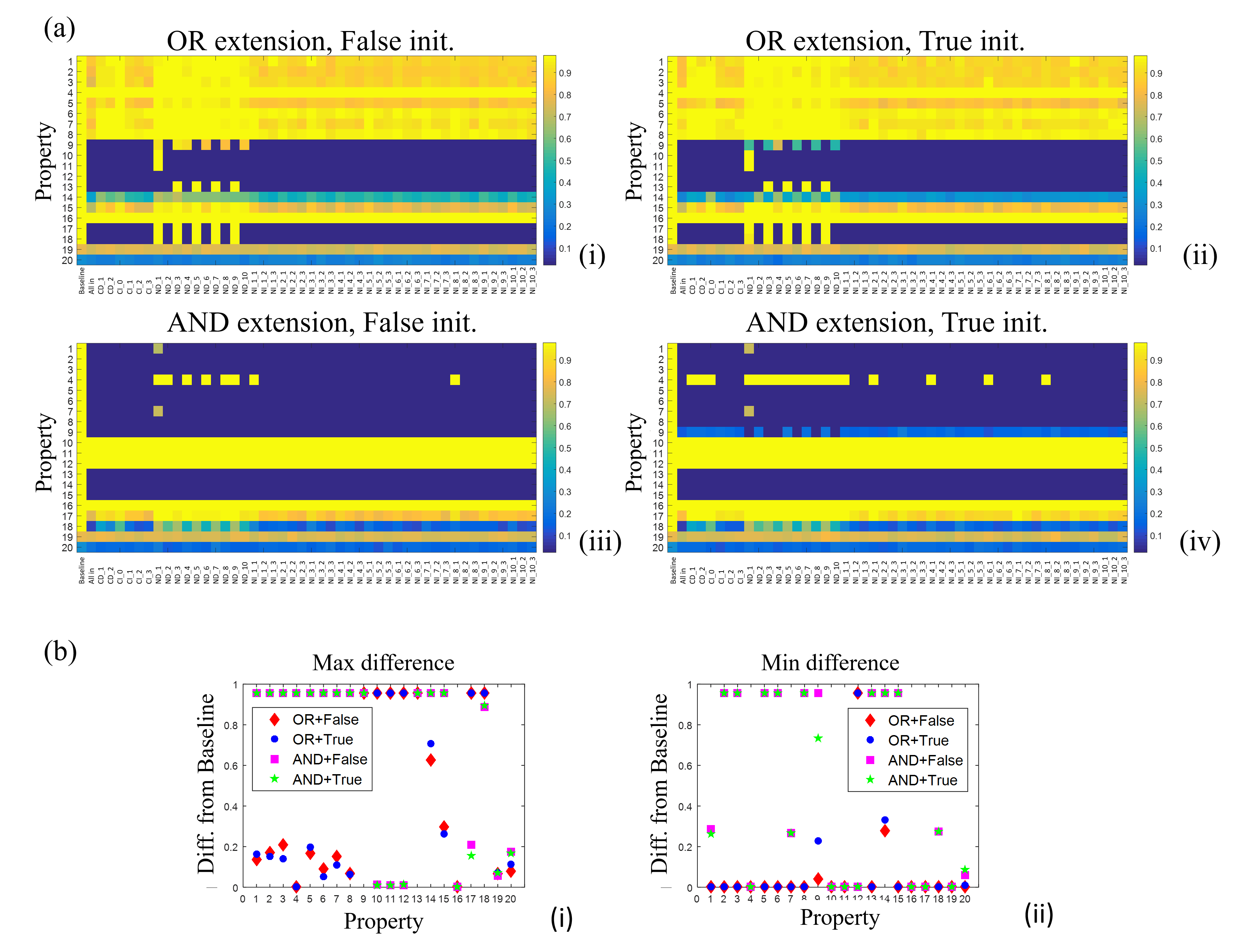}
    \caption{(a) Results of statistical model checking of 20 properties in 68 different models. Each entity in x-axis is a model, and each row is the estimated probability for the corresponding property. (b) The Max and min difference from the baseline model of each property.} 
    \label{fig:properties}
\end{figure}

In Fig. \ref{fig:properties}(a), four different sections can be observed, and each section corresponds to one of the extension methods. Each method has its unique feature. For example, the ND method only includes relationships relevant to one layer, and this makes the number of new elements added to the model significantly smaller than other methods. Also, the light blue nodes indicate the number of newly added elements that are in a newly formed pathway. Since CD and ND do not include indirect parent interactions, we can see that the number of elements in new pathway is $0$. While in CI and NI, we can tell that indirect interactions are included. The numbers within one method show higher similarity, but we can still observe some patterns. For example, the cumulative parent-set methods, CD and CI show an increase in the number of new nodes when more layers are considered. Furthermore, since NI has cumulative parents when they finish the noncumulative part, they also experience an increase when the step of noncumulative part is fixed. The numbers saturate at around $600$, which is due to the limited size of baseline model and extensions we have. This is also the reason we choose to perform the cumulative approach for at most $3$ steps. 

In general, choosing the method to extend the model depends on the scenario a user is interested in. For example, if the focus is on the regulation of a specific element, one can track down each layer of parents using ND, and see the change of the model after modifying that specific layer. On the other hand, if the goal is to include as many new stimuli as possible with a fewer number of layers, cumulative methods such as CI or CD will fit better. We selected 20 elements as part of the base layer, since these elements appear in properties that we are testing, leading to relatively large base layer given the size of the baseline model. Therefore, by incorporating elements related to more than one layer, we capture almost all extensions related to the baseline model. Thus, the 'All In' method, which adds all extension interactions to the baseline model at once, does not change the counts shown in Fig. \ref{fig:properties}(a), when compared to many cases of CD, CI and NI methods.

\subsection{Impact of model extension on system properties}

Fig. \ref{fig:properties}(a) shows the results of testing $48$ models (baseline $+$ All-In $+$ $46$ extended models) with different extension method ($AND$/$OR$) and different initialization of the newly added elements ($True$/$False$) against the 20 properties in Table \ref{tab:properties}. The values displayed are the estimated probabilities of each property. Just like the basic numbers of each model, different extension methods lead to different results of the properties. For example, we can see that the results from ND are different from other methods. The reason is that each ND method only deals with one layer at a time, and it will not insert new edges between elements mentioned in the properties. This leads to a more conservative extension. Also, for example, there are differences between OR-based ND models in properties 9 to 13 or property 4 in AND-based ND models, which are related to Inhibition of tumor suppressors and Autophagy in PCCs. By comparing the extension interactions added to those models, we found that the EGF (Epidermal Growth Factor) pathway plays the most important role. The p21 (regulator of cell cycle progression) pathway also influences the difference.

If we compare the models with different initialization of newly added nodes, we can see the results are actually quite similar. This means that the model is mostly influenced by the input elements in the baseline model, and to some degree, it emphasizes the robustness of the original model. On the other hand, if we compare extending the models with OR operations and those with AND operations, there is a huge difference. But the interesting part is that the behavior of models with the two types of extensions is opposite. They behave similarly only in properties 9, 13, 16, 19 and 20, while differently in all other 15 properties. This shows a drastic difference between AND-based and OR-based extension, and can be further designed according to the property we want to fit. Fig. \ref{fig:properties}(b) shows the maximum / minimum difference compared to baseline that each model can achieve for each property. If a property probability is low in both max and min difference, it is relatively conservative to the extension interaction. An example is property 16, which depicts the relationship between p53 and Apoptosis. On the other hand, if a property probability is high in both max and min difference, it is a property susceptible to changes via extensions.

\section{Discussion}

The framework we describe here, although designed to extend an existing baseline model, can also be used to search for pathways or interactions that are vital to certain functions, and to suggest targets for drug development. For example, using the ND models and statistical model checker, we can study closely how each layer of elements influences the elements we are interested in. Then, we can pin-point the models that satisfy several properties that we desire, and we should be able to identify a few candidates that play important roles in the regulation. Or, by using NI method, we can further observe whether there is actually an upstream network that controls the behavior of the elements. This gives us a deeper understanding of the network and helps us in further model development.

One of our next steps is to improve the approach to incorporate new elements into logical rules. In this work we naively incorporate those rules using $OR$ and $AND$ operation. However, in reality the mutual relationships between the regulators are not necessary an $AND$ or $OR$ relationship. For example, a ligand and a receptor induce further response if they both exist, and there is another unrelated element activating the same target. This results in a format $A=B*C+D$. We are not able to capture this since the automated reader does not output this information, but from online databases such as UniProt\cite{uniprot2014uniprot}, we are still able to gather pieces of knowledge about the true interaction between regulators. Also, the automated reader does not output the location of the interaction. For example, two types of cells, PCCs and PSCs, are in our baseline model, but we only extend the interactions to PCCs. More information of the location can also help us refine the extension method. As a future work, incorporating the on-line database should give us a more accurate extension of the model. But in the long run, if the automated reader can take into account these features, we should be able to construct a better model more easily. Finally, aside from extensions, the automated reader provides us with contradictions. In this work we ignore this kind of relationship and assume absolute correctness of interaction in the baseline model, but the contradictions serve as a great starting point to examine the validity of the baseline model, as well as to point to further improvements of reading engines.

\section{Conclusion}
We propose a framework that utilizes published work to collect extensions for existing models, and then analyzes these extensions using stochastic simulation and statistical model checking. With biological properties being formulated as temporal logic, model checker can use the trace generated by the simulator to estimate the probability that a certain property holds. This gives us an efficient approach (speed-up from decades to hours) to re-use previously published results and observations for the purpose of conducting hundreds of \textit{in silico} experiments with different setups (models). Our methods and the framework that we have developed comprise a promising new approach to rapidly and comprehensively utilize published work for an increased understanding of biological systems, in order to identify new therapeutic targets for the design and improvement of disease treatments.

\section*{Acknowledgement}

We would like to thank Mihai Surdeanu (REACH team) and Hans Chalupsky (RUBICON team) for providing output of their reading and assembly engines, and Michael Lotze for his guidance in studying cancer microenvironment. This work is supported by DARPA award W911NF-14-1-0422.

\bibliographystyle{splncs03}

\end{document}